# Properties of Two-Dimensional Silicon grown on Graphene Substrate

R. Zhou[1], L. C. Lew Yan Voon[1,2,3], Y. Zhuang[1]

[1]Department of Electrical Engineering, Wright State University, Fairborn, OH 45435, USA
[2]Department of Physics, Wright State University, Fairborn, OH 45435, USA
[3]School of Science and Mathematics, The Citadel, Charleston, SC 29409, USA

The structure and electrical properties of a two-dimensional (2D) sheet of silicon on a graphene substrate are studied using first-principles calculations. A new corrugated rectangular structure of silicon is proposed to be the most energetically favorable structure. The shifting of the Fermi energy level indicates self-doping. Calculation of electron density shows a weak coupling between the silicon layer and graphene substrate. The 2D silicon sheet turns to be metallic and has a much higher value of transmission efficiency (TE) than the underlying graphene substrate.

Inspired by the success of graphene[1-4], various two-dimensional (2D) structures on different substrates have been proposed. Among others, a new allotropic form of silicon, coined "silicene" by Guzmán-Verri and Lew Yan Voon in 2007[5], was shown to exhibit similar properties with graphene of a zero band gap and a Dirac cone shape at the K point. Earlier, free-standing silicon had already been predicted to have a buckled hexagonal aromatic stage.[6, 7] With a Fermi velocity slightly smaller than graphene, silicene nevertheless hold higher mobility than many other materials.[5] Recently, high-performance field-effect transistors have been proposed based upon silicene nanoribbons.[8] Thus, silicene can be expected to have all of the exciting properties of graphene, with the added benefit that its fabrication and processing would be compatible with the existing silicon-based semiconductor platform, lowering the barrier to mass fabrication of silicene electronic devices. One of the main obstacles is to grown silicene with controllable electric transport properties. It is well know that substrate could impose great impacts on the structural and electric properties of the films grown on it. To date, successful growth of silicene has only been reported on metallic substrates of silver (Ag)[9-15] and zirconium diboride ($ZrB_2$).[16] Exploration of silicene on various substrates is of key importance to tailor its structural and electrical properties.

In 2010, a periodic bi-layered silicon/graphene superlattice grown on graphene substrate was explored.[17] The structure was proposed mainly based on the theoretical finding that the bond length of graphitic Si is 2.35 Å,[18] close to graphene's lattice constant of 2.46 Å.[17] As a result, silicon atoms should sit well at the H site (Hollow of hexagonal shape, see Fig. 1(a)), forming a flat triangular stage to best match the graphene structure. However, the flat formation of silicon is not guaranteed to be the most energetically favorable one. In this work, density functional theory (DFT) is employed to optimize the atomic structure of a 2D silicon monolayer on graphene substrate by looking for the relaxed structure with the lowest total energy. A corrugated rectangular structure is obtained and verified to be the most energetically favorable structure. Energy band diagram, distribution of electron density, density of states (DOS), transmission efficiency (TE) and current-voltage (I-V) of the structure are computed, from which the 2D silicon monolayer is found to be metallic and has a weak coupling with the graphene substrate.

DFT calculations (Atomistix Toolkit (ATK)) were carried out using norm-conserving pseudopotentials coupled with non-equilibrium Green's function (NEGF).[18-20] The exchange-correlation function is approximated using the local density approximation with the Perdew-Zunger parametrization.[21] The wave function is expanded in terms of a local atomic orbital basis set and a double-zeta polarized basis set is used for all the calculations. Spin-orbit coupling terms are not included in this study. The Monkhorst-Pack grid k-point sampling is chosen to be 21x21x1. "1" is chosen for the vertical direction, because of the absence of periodicity. "21" is tested to be sufficient for convergence. The vertical size of the unit cell is chosen to be 30 Å to avoid interlayer interaction.

With a force tolerance of 0.001 eV/Å in most cases, the graphene substrate is set to be fully constrained, referring to a rigid substrate. The periodic boundary condition is used in all directions when solving the Poisson equation self-consistently by the fast Fourier transform solver. All calculations are carried out when the electron temperature is at room temperature, and with a density mesh cut-off of 75 Hartree.

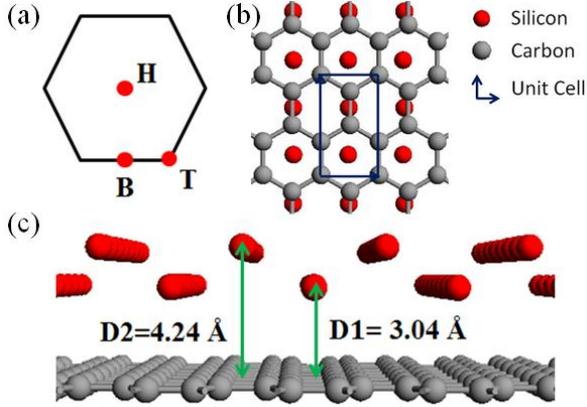

FIG. 1. (a) Hollow, Bridge and Top sites on a hexagonal cell. (b) 3-D top view of Rec-Zig structure. Red balls represent silicon atoms, forming into rectangular lattice. Half of silicon stands on B site, the other half stands on H site. Blue balls represent carbon atoms. Blue rectangular arrows show the unit cell, without showing vertical side. (c) 3-D side view of Rec-Zig structure. The distance from Bridge site silicon atoms to graphene D1 is 3.04Å, while the distance from Hollow site silicon to graphene D2 is 4.24 Å. Zigzag shape can be observed from this side view.

The initial model contains a 2 x 1 graphene hexagonal Bravais lattice unit cell with two silicon atoms and four carbon atoms. The graphene substrate of a lattice constant 2.46 Å is constrained, while relaxation is allowed for silicon atoms along all directions. The silicon atoms relax to a buckled stage, with half of the atoms located on the B site (Bridge of hexagonal shape, see Fig. 1(a)), and the other half located on the H site. We further constructed an orthorhombic super cell with two silicon atoms and four carbon atoms, as shown in Fig. 1(b). From this top view, the rectangular formation of Si atoms can be observed (Fig. 1(b)). If viewing by the side, a zigzag shape can be observed. B site silicon atoms have a distance of 3.02 Å above the graphene layer, while H site silicon atoms are 4.24 Å away (see Fig. 1(c)). The Si-Si bond length is about 2.45 Å, similar to graphene's lattice constant value, but a slightly larger than Si bond length of a diamond structure.

Since it consists of rectangular lattice and zigzag lattice (see Figs. 1(b) and 1(c)), this structure is labeled as "Rec-Zig".

To verify the stability and reproducibility, simulations were performed with various initial positions of silicon atoms (Fig. 2) with respect to the graphene substrate. Orthorhombic super cell is used in this calculation. Due to the symmetry, silicon atoms will take T (top of a carbon atom), H, or B sites. Figure 2 shows all the 11 possible combinations of the two silicon atoms in an orthorhombic unit cell. The simulation results turn out that no matter which initial position was chosen for the silicon atoms, the final relaxed structure always approaches to the proposed Rec-Zig structure (i.e. Fig. 2(k)).

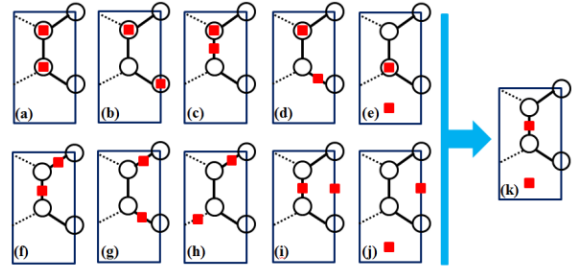

FIG. 2. 11 different possible combinations of silicon atoms grown on graphene substrate. Each unit cell include four carbon atoms (black open circles) and two silicon atoms (red squares). (a) and (b) are possible T-T structures; (c) and (d) are possible B-T structures; (e) is the only possible H-T structure; (f), (g) and (h) are the possible B-B structures; (i) and (k) are the possible H-B structures; and (j) is the only possible H-H structure. Among these combinations, (k) is the Rec-Zig structure, and (j) is the FHB structure. All possible structural combinations relax to the (k) structure, namely the Rec-Zig structure.

For comparison, a structure of flat hexagonal Bravais lattice (FHB) is calculated in this letter for the following discussion (Fig. 2(j)). Different from the three-dimensional (3D) superlattice model[17], FHB is a 2D structure. With a force tolerance of 0.05 eV/Å, the spacing between the silicon layer and graphene layer is relaxed to be 3.37 Å, about 10% smaller than the silicon/graphene superlattice calculated by Zhang and Tsu.[17] Compared to the total energy of the Rec-Zig structure $E_{\text{Rec-Zig}}^{tot} = -164.6244 eV/atom$, the FHB is about 0.05 eV/atom higher, showing the Rec-Zig structure is more energetically favorable. This may be explained by a study of Aktürk *et al.*[21], in which they found Si atoms are bound with a most significant energy on the B site instead of H and T on graphene.

The energy band diagrams of the Rec-Zig and FHB structures are shown in Fig. 3. The Fermi energy level of the Rec-Zig structure is found to move 0.48 eV upward above the Dirac point (Figs. 3(c), 3(d), 3(e), and 3(f)), which is about a 50% larger than that in FHB ~0.32eV (Figs. 3(a) and 3(b)). This indicates self-doping occurred, i.e., electrons transferring from the silicon layer to the graphene substrate.

The density of states (DOS) of the Rec-Zig and the FHB structures are calculated and shown in Fig. 4. Near the Fermi energy level, the DOS for both structures are very similar. However, the TEs obtained using non-equilibrium Green's function (NEGF) theory, shown in Fig. 5(a), are significantly different from each other: The corrugated structure Rec-Zig reaches the minimum of the TEs close to the Fermi energy level, while the TEs of the FHB structure are continuously decreasing and crosses the Fermi energy level. To gain a more in-depth understanding, calculations of TEs on de-coupled single silicon layer and pure graphene substrate were carried out (Figs. 5(b) and 5(c)). The TEs of the underlying graphene from both the flat and corrugate structures are the same, while the TEs of silicon in both structures are much higher than those of graphene. This indicates that silicon layer contributes most in the transmission.

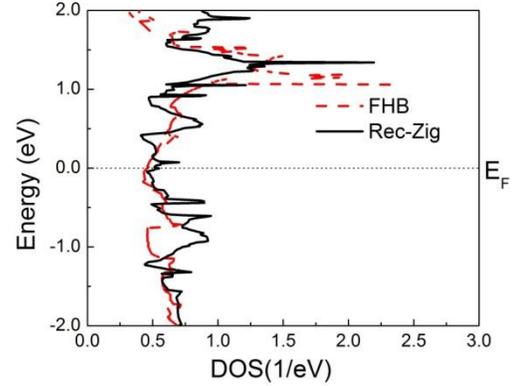

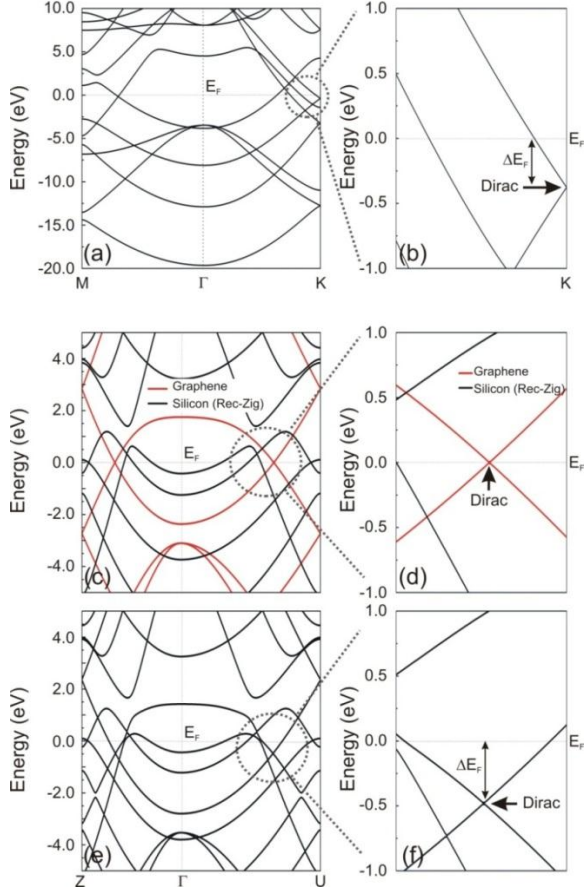

FIG. 3. The band diagrams of FHB and Rec-Zig structure, with the Fermi level shifting. (a) FHB structure's band diagram. M, Γ, K are the three symmetrical points in the hexagonal Bravais lattice Brillouin zone. (b) The enlarged figure around Dirac point, showing Fermi level moves $\Delta E_F = 0.32$ eV above Dirac point, which located on K point. (c) Rec-Zig structure's band diagram, showing silicon and carbon layers separatedly, without coupling. Z, Γ, U are the symmetrical points in the simple orthorhombic Bravais lattice Brillouin zone. Red bands are for Graphene layer, while black bands for silicon layer. (d) Enlarged figure of Dirac point, showing the Dirac point at Fermi level without coupling from silicon. Here the Dirac point located in between two symmetrical points in Brillouin zone. (e) Rec-Zig structure's band diagram, with coupling. (f) Enlarged figure of Dirac point, showing the coupling makes Fermi level move $\Delta E_F = 0.48$ eV above Dirac point.

FIG. 4. Density of States and I-V curve of the FHB and Rec-Zig structure. Inset shows the current-voltage relationship (I-V curve) of FHB and Rec-Zig structure.

The current-voltage curves are shown in Fig. 6, which are computed from

$$I(V_b) = \frac{2e}{h}\int_{-\infty}^{+\infty}\{T(E,V_b)[f(E-\mu_L)-f(E-\mu_R)]\}dE, \quad (1)$$

where $T$ is the transmission probability at a given bias $V_b$, $f(E)$ is the Fermi-Dirac distribution function, and μ is the electrochemical potentials of the left (L) and right (R) electrodes. Eq. (1) is the Landauer-Büttiker expression for ballistic transport.[24] Obviously both the Rec-Zig and the FHB structures exhibit linear dependency between current and voltage, indicating metallic features of both structures. This coincides with the absence of the energy band gap in Fig. 3. In addition, due to the higher value of the TEs, the flat structure shows a higher electrical conductivity than the corrugated structure.

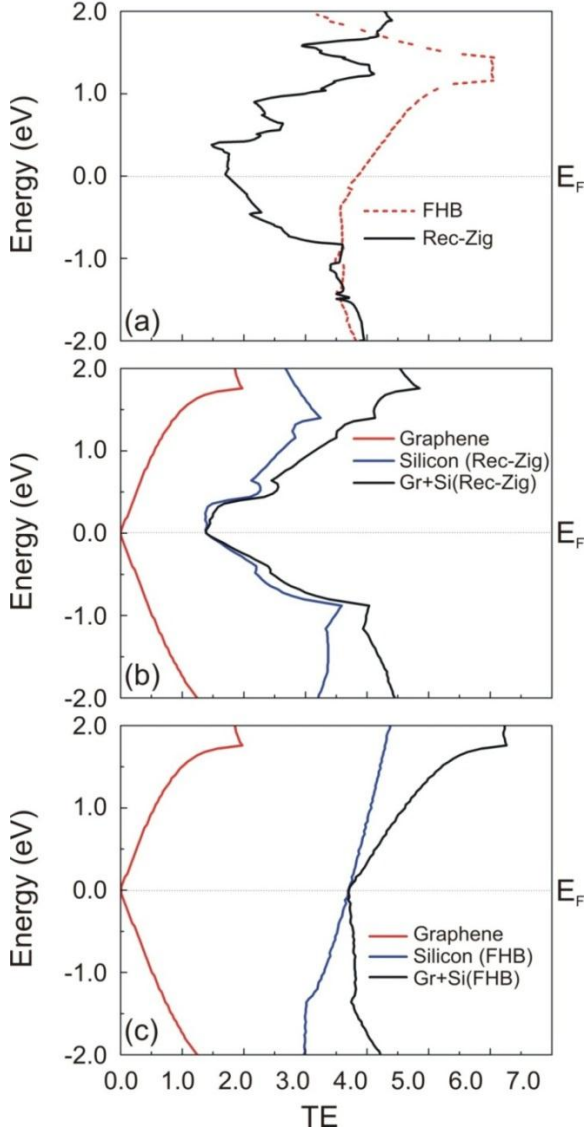

FIG. 5. The transmission efficiencies of two structures and their separate components. (a) Transmission efficiencies of the FHB and Rec-Zig structure. (b) Transmission efficiencies of separate graphene and silicon components of Rec-Zig structure and their sum value. (c) Transmission efficiencies of separate graphene and silicon components of the FHB structure and their sum value.

To further understand the transport properties, distribution of electron density was calculated and shown in Fig. 7. Strong couplings between C-C and Si-Si were observed, while the Si-C coupling is rather weak. This is in agreement with the large Si-C spacing ~ 3.04 Å of the Rec-Zig structure. The binding energy between Si layer and Graphene Layer in one unit cell is defined by $E_{binding}=(E_{SiLayer}+E_{Graphene}-E_{Rec-Zig})/N$, where $E_{SiLayer}$ is the total energy of the corrugated rectangular Si layer separate from grapheme, $E_{Graphene}$ is the total energy of graphene atoms, $E_{Rec-Zig}$ is the total energy of the two layers combined together, N is the total number of atoms of the unit cell. This binding energy of Rec-Zig is calculated to be 0.05 eV/Atom, which is in the same order as the Van der Waals binding between graphene sheets when their distance is about 3A: 0.037 eV/Atom.[25] This suggests that Van de Waals force might be co-existent. In addition, the smaller distance of about 3 Å between the closest Si and the graphene layer, similar to the C-C interlayer distance (~ 3.3 Å)[25] for bilayer graphene in spite of the larger size of Si atoms, indicate that the Si-C interaction is stronger than the interlayer coupling for bilayer graphene. Hence, the Si-C coupling is intermediate between the strong covalent one and the weak Van der Waals one. Further analysis including Van der Waals interactions is currently under developing.

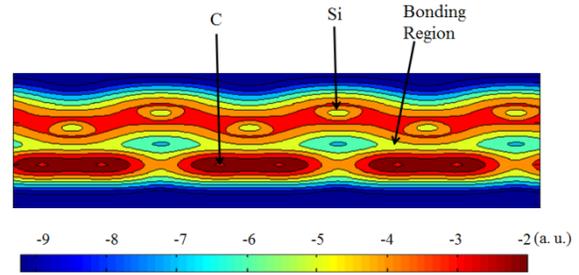

FIG. 7. Electron Density of Rec-Zig structure. The zigzag black line is Si layer, the horizontal red shape indicates graphene layer. Weak bonding region (yellow) between the layers is observed. This result is plotted by logarithmic scale.

In conclusion, silicon-on-graphene structures are investigated using density-functional calculations. A new energetically favorable structure Rec-Zig was found of having a 1.22 Å buckling vertically within the silicon layer. Half of the silicon atoms are located on the B site, and the other half located on the H site. The Rec-Zig structure shows a more pronounced self-doping, linear dependency of current on voltage, weak Si-C coupling, and a lower ballistic conductance than the FHB structure. We believe the results should motivate the growth of large-area silicene sheets on graphene.

**Acknowledgements**

R. Z. and Y. Z. would like to acknowledge the support of the Research Challenge Award by Wright State University.